\documentclass[12pt]{article}
\usepackage{amsmath}
\usepackage{amssymb}
\usepackage{amscd}

\newcommand{\F}{\mathbb F}

\begin{document}

\begin{center}
\large\bf
A Diffie-Hellman key exchange protocol using\\ matrices over non commutative rings \\

\end{center}
\begin{center}
\small Mohammad Eftekhari\\
LAMFA, universit\'e  de Picardie-Jules Verne\\33 rue Saint-Leu 80039 Amiens France

email:mohamed.eftekhari@u-picardie.fr\\
\end{center}

\begin{center}
{\large\bf
Abstract}
\end{center}

{\it We consider a key exchange procedure whose security is based on the difficulty of computing discrete logarithms in a group, and
 where  exponentiation is hidden by a conjugation. We give a platform-dependent cryptanalysis  of this protocol.
 Finally, to take full advantage of this procedure, we propose a group of  matrices over  a  noncommutative ring  as platform group.\\

\vfill
}\\

{\bf\large 1.Introduction} \\

The Diffie-Hellman  key agreement  protocol is the first published practical solution to the key distribution problem, allowing two parties that have never met to exchange a secret key over an open channel. It uses the cyclic group $\F_q^*$, where $\F_q$ is the finite field with 
$q$ elements. The security of this protocol is based on the difficulty of computing discrete logarithms in the group $\F_q^*$.\\
 There are several algorithms  for computing discrete logarithms, some of them are subexponential  when  applied
to $\F_q^*$.\\

 It is  important to search for easily implementable groups, for which the DL problem is hard and there is
no subexponential time algorithm for computing DL. The group of points over $\F_q$ of an elliptic curve is such a group.\\
So keeping in mind the above remarks and the fact that $\F_q^*=GL_1(\F_q)$, one can  wonder whether  the group $GL_2(\F_q)$ of 
two-by-two invertible
matrices or more generally  the group  $GL_n(\F_q)$, which  admit  a``natural'' normal form,  can be used for  a  Diffie-Hellman protocol and  whether  there is some advantage  in  using them.
\\

{\bf Remark 1.1}
 Let  us  fix a matrix $X\in GL_n(\F_q)$. Knowing $X$ and a power $X^a$, is it easy to find $a$?
The first point is that  knowing $X$, one can compute $det(X)\in\F_q^*$ (the determinant of $X$), and also $det(X^a)=(det(X))^a$.
 In  this way, the DL problem in matrix groups  reduces  to  the  DL problem in $\F_q^*$.\\

One can avoid this difficulty by  choosing  a matrix $X$ such that $det(X)=1$, but then by computing eigenvalues of $X$ and of
$X^a$ ( possibly  in an extension of the base field), and using the fact that the  latter  are the  former in the power of ~$a$,  one reduces
 once again the DL problem
to the one in some extension of $\F_q^*$.\\

So there is no advantage of considering the DL problem in the group of invertible matrices over a finite field, and more
generally over a finite commutative ring.\\

We wish to mention that the group of matrices  over a finite field as above was first  proposed as  a  platform group for Diffie-Hellman key exchange
in [12], and was cryptanalysed using eigenvalues and Jordan form in [10]. Note that in this proposition the noncommutative structure of $ GL_n(\F_q)$
is not used.\\

In [2], a protocol using noncommutative (semi) groups in cryptography was proposed. A platform using braid groups and the same idea was proposed
in [9]. Also another platform  using matrix algebra was discussed in [16].
 The protocol we use in section 3  is based on the same  idea. 
It uses conjugation and exponentiation together for its security. A platform for this protocol  using braid groups was first proposed in [14] and another one using
an $\F_q$-algebra in [11]. We  shall  give a cryptanaly\-sis of these two platforms in section 3,  by reducing the problem to the discrete logarithm problem over some finite field .\\

The  semigroup   of matrices over a commutative ring  was  considered in [8] for an authentication
protocol, but its security is  based on the difficulty of  the  conjugacy search problem and not on the discrete logarithm one. In fact the authors consider matrices over a somehow complicated ring,
namely the ring of N-truncated  polynomials in k variables to make the conjugacy search problem infeasible.\\

To avoid the reduction of DL problem to the one over finite fields mentioned in the above remark,  which stems from the  special features of  (semi)-group matrices over finite fields (namely determinant
and properties of eigenvalues), 
we can consider matrices over noncommutative finite rings.  Group algebras $\F_q[G]$, where $G$
is a noncommutative finite group are examples of such rings. The simplest example of such group algebras  is the group algebra of the group of  permutations of three
elements, which is easily implementable. We can then consider  two-by-two invertible matrices over such a group algebra. In the next section,
one  considers  matrix groups over noncommutative rings and  investigate whether  the  previously mentioned  reduction (remark 1.1)  in the case
of DL problem in the group matrices over finite fields can happen or not.\vfill

{\large\bf 2. Quasideterminants, noncommutative determinants, eigenvalues...}\\

Since the invention of quaternions, there has been attempts to define a notion of determinant of a matrix with
noncommutative entries.  Here  one can  mention  great names such as Cayley, Study, Moore, Wedderburn, Heyting and Richardson, Ore, Dieudonn\'e, Berezin,  who all considered  such noncommutative determinants. In most of the cases, these noncommutative
determinants are rational functions of the entries. The most recent and most general attempt (1991) is due to
I.~Gelfand and Retakh.  It proved to be  very effective in many areas of noncommutative algebra. In what follows we recall some definitions and results from  [4], [5], [6], [7].  See also [15], for a generalization of Dieudonn\'e determinant.\\

Given a square matrix $A$ of size $n$, with entries in a noncommutative ring $R$, we note $A^{ij}$ the matrix obtained
from $A$ by deleting the  $i$th  row and  the $j$th  column. We also note by $r_i^j$ the  $i$th  row of $A$ with  $j$th 
position excluded, and by $c_i^j$ the  $j$th  column of $A$ with the  $i$th  position excluded. For each position $(i,j)$, the  quasideterminant of $ A$ is defined by
$|A|_{ij}:=a_{ij}-r_i^j(A^{ij})^{-1}c_i^j$. We have  $|A|_{ij}\in R$ and, of course, this quasideterminant exists if the $(n-1)$-by-$(n-1)$ matrix $A^{ij}$
is invertible. So, for a matrix of size $n$, there are $n^2$ quasideterminants. \\

Example: $n=2$, $A=\left(\begin{array}{cc}
a_{11}&a_{12}\\
a_{21}&a_{22}
\end{array}\right)$\\
$|A|_{11}=a_{11}-a_{12}a_{22}^{-1}a_{21}$\\
$|A|_{12}=a_{12}-a_{11}a_{21}^{-1}a_{22}$\\
$|A|_{21}=a_{21}-a_{22}a_{12}^{-1}a_{11}$\\
$|A|_{22}=a_{22}-a_{21}a_{11}^{-1}a_{12}$\\

{\bf Remark 2.1}: Even in the commutative case, a quasideterminant is equal  not  to a determinant, but to the ratio of two determinants,
namely, $|A|_{ij}=(-1)^{i+j} {det(A)\over  det(A^{ij})}$.\\

Using quasideterminants, one defines a noncommutative determinant which gives the determinant (modulo a sign)
in the commutative case:\\
Let $I=\{i_1,i_2,...,i_n\}$ and $J=\{j_1,j_2,...,j_n\}$  be  two orderings of the set $\{1,2,3,...,n\}$. Note by
$A^{i_1i_2...i_k,j_1j_2...j_k}$ the matrix obtained from $A$ by deleting the lines $i_1,i_2,...,i_k$ and the columns
$j_1,j_2,...,j_k$. Then one defines the noncommutative determinant of the $n$-by-$n$ matrix $A$ by:\\
$D_{I,J}(A):=|A|_{i_1,j_1}|A^{i_1,j_1}|_{i_2,j_2}
|A^{i_1i_2,j_1j_2}|_{i_3,j_3}...|A^{i_1i_2...i_{n-1},j_1,j_2...j_{n-1}}|_{i_n,j_n}$. \\
Example: For a two-by-two matrix
$A=\left(\begin{array}{cc}
a_{11}&a_{12}\\
a_{21}&a_{22}
\end{array}\right)$, we  find 
$I=J=\{1,2\}$ and $D_{I,J}=(a_{11}-a_{12}a_{22}^{-1}a_{21})a_{22}=a_{11}a_{22}-a_{12}a_{22}^{-1}a_{21}a_{22}$

Using this noncommutative determinant one can recover some of the  previously considered notions such as the  Dieudonn\'e determinant.\\

There is  still  another definition of  a  noncommutative determinant [4], motivated by representation theory, and giving
the determinant in the commutative case. This noncommutative
determinant is an elementary  symmetric  function of the noncommutative eigenvalues of $A$.
We  do not  give this definition here.\\

 To summarize,  there is an active area of noncommutative algebra dealing with noncommutative determinants, noncommutative 
eigenvalues, ... From our cryptographic point of view, we  only need  to make sure that there is no formula
reducing the DL problem in the group of matrices with noncommutative entries to the DL problem in the ring of coefficients.
To the best of our knowledge, there is no way to relate the determinant of a matrix or its eigenvalues to the corresponding
determinant and eigenvalues of a power of this matrix in the noncommutative case.\\

{\large\bf 3. A Diffie-Hellman key exchange protocol}\\

We consider the following protocol, which is based on the general idea of [2]. The  platform  proposed in [9] using braid groups is based on the same idea; in the latter  case,  the security is based
on the conjugacy search problem, whereas in the following,  one uses the discrete logarithm and the conjugacy search problem together.\\
Suppose $G$ is a noncommutative group and $H_1$ and $H_2$  two  subgroups of $G$ such that every element of $H_1$ commutes with
every element of $H_2$.\\
 Here  $G,H_1,H_2$ and an element $X\in G$ of some  high order $n$ will be public data. Alice and Bob  will  use these data to exchange a key.\\

Alice selects at random a secret integer $a\in\{2,3,...,n-1\}$ and a secret element $T\in H_1$ ($TX\not = XT$); she computes $TX^aT^{-1}$ and
sends it to Bob.\\

Bob selects at random  a secret integer $b\in\{2,3,...,n-1\}$ and a secret element $T'\in H_2$ ($T'X\not =XT'$); he computes $T'X^bT'^{-1}$
and sends it to Alice.\\

Alice computes $(T'X^bT'^{-1})^a=T'X^{ab}T'^{-1}$; then she conjugates it by her secret element $T$ to
obtain $TT'X^{ab}T'^{-1}T^{-1}$.\\

Bob computes $(TX^aT^{-1})^b=TX^{ab}T^{-1}$ and he conjugates it by his secret element $T'$ to obtain
$T'TX^{ab}T^{-1}T'^{-1}$ which is the same as what Alice obtained due to the commutativity  $TT'=T'T$.\\

 We  immediately  see  that the choice of  a  matrix group over  a  finite field (and to some extent over  a  commutative ring)  as a platform group
for this protocol is not a good one. In fact, Remark~1.1 in the introduction  about  the reduction of the DL problem from
matrix groups to the same problem over some extension of the base field remains valid. Let $\lambda$ be an eigenvalue of   $TX^aT^{-1}$.  One   has
$\det( TX^aT^{-1}-\lambda id)=0$, so $\det(T(X^a-\lambda id)T^{-1})=0$. Then, by the multiplicative property of determinant, we get 
$\det(X^a-\lambda id)=0$ and $\lambda$ is an eigenvalue of $X^a$ and is equal to some eigenvalue of $X$ to the power $a$.\\
So  choosing a  matrix
group over  a  finite field as  a  platform group  offers no advantage. 
Furthermore, taking the underlying multiplicative group of an algebra as platform group does not provide any advantage  either, as using  representation theory one can reduce the problem to the one over matrices and then to the discrete logarithm over some finite field.\\

This protocol was first used  in [14] in the context of braid groups. In the paper the authors consider a modified irreducible Burau type representation of a braid group  and
apply this protocol at the representation level to the matrices over some finite field. By what we said previously this is not a good choice and can be reduced to the DL problem over some extension of the field.\\
The same protocol was used in [11], by taking as the platform group the multiplicative group of a noncommutative algebra of dimension four over a finite field. By taking the regular representation of this algebra we can transfer the scheme to the level of matrices and then reduce it to the DL problem
in some extension of the finite field.\\
In [13] this protocol is  implemented  as a software for smartphones  using  ($5 \times 5$) matrix  groups over a finite field, and its performance  is  compared to other implementations
using finite fields or elliptic curves. The result of this comparison is that this protocol is largely more performant than  those  using finite fields or elliptic curves. As  mentioned  before,  due to  the reduction to the case of discrete logarithm over a finite field, the performance of this protocol using
matrix groups over a finite field must not be so different from the one over a finite field.\\
So, to take the best advantage of this protocol, we propose  to choose as a  platform group the group of matrices over  a  noncommutative rings, namely we
consider two by two matrices over the group algebra of the symetric group $S_3$, which we denote by $G=GL_2(\F_q[S_3])$ where $S_3$ is the
group of permutations of three elements.  Here  $X$ will be an element of $GL_2(\F_q[S_3])$ and  we fix 
$$ H =  H_1=H_2=\Big\{\left(\begin{array}{cc}
x&y\\
y&x
\end{array}\right) \in GL_2(\F_q) \mid x\in\F_q,y\in\F_q, x^2-y^2\not= 0\Big\}, $$
which is a commutative
subgroup of $GL_2(\F_q[S_3])$. In fact $H$ is a maximal torus of  $GL_2(\F_q)$. \\

{\large\bf 3.1  ElGamal  encryption}\\

Suppose that Alice is the owner of the public key data, $GL_2(\F_q[G])$, $X\in GL_2(\F_q[G])$
of order $n$ and  $H=H_1=H_2$ as above.
Suppose also that Alice has selected a secret integer $a$ and a secret matrix $T\in H$, and made
$TX^aT^{-1}$  public.  Bob can encrypt a message $M$ intended for Alice, as follows:\\
Bob selects a random integer $b\in\{2,3,..,n-2\}$, and a matrix $T'\in H$;\\ he computes $TT'X^{ab}T^{-1}T'^{-1}$
as explained in the precedent section.\\
Bob determines a symmetric encryption key $t$ based on $TT'X^{ab}T^{-1}T'^{-1}$ (in  a way he agreed upon with Alice).\\
Bob uses an agreed upon symmetric encryption method with key $t$ to encrypt $M$, resulting in the encryption $E$.\\
Bob sends $(T'X^bT'^{-1},E)$ to Alice.\\
Receiving these data, Alice computes $TT'X^{ab}T^{-1}T'^{-1}$, as in the previous section; she derives from
this the symmetric encryption key $t$; she uses the agreed upon symmetric encryption method with key $t$ to
decrypt $E$, and finds $M$.
\vfill

{\bf Remark 3.1.1} The  ElGamal  encryption as explained above is an hybrid version of ElGamal's encryption. In the
textbook ElGamal encryption, we can take the message $M\in GL_2(\F_q[G])$:\\
Bob sends to Alice
$(T'X^bT'^{-1},TT'X^{ab}T^{-1}T'^{-1}M)$. \\
Alice computes $(TT'X^{ab}T^{-1}T'^{-1})^{-1}$ and by
multiplying at the left with the second data, she finds~$M$. See also [1].\\

{\large\bf 4. Choice of parameters and security}\\

 Owing to  the similarity between the  protocol we use and  the one proposed  in the
context of braid groups [9],  one may ask if the same kind of attacks as in the braid groups can be applied in
our context.\\

We remind that the security of braid-based cryptography  relies on the difficulty of the conjugacy search  problem.  The problem is  as follows:  Knowing an element $X$ and a conjugate $TXT^{-1}$,
is it easy to find $T$?  In other words,  we know an element and some conjugate of it and one tries to find  a conjugating  element  $T$.
One of the main attacks against these procedures is to search~$T$ not in the  whole  conjugacy class of $X$, but in
some  characteristic  part of it. The second kind of attack is to use some probabilistic research in the conjugacy class of $X$. The third one is to use linear representations of braid groups to reduce the problem to the one in  a  matrix group,
which is easy to solve. See [3] for details.\\

The main difference  between our approach and those  using braid groups is  that, in our case,  $X$ is publicly known, but the conjugacy class which is involved is that of $X^a$, which is not known, so all the above attacks are  useless  in our case.\\

As we mentioned before (section 2), specific features of the group of invertible matrices with noncommutative entries 
 cannot  be used to attack our protocol.\\

 As for  the existing algorithms computing discret logarithms, such as ``Baby Step, Giant Step",  or the  Pollard rho algorithm,  they  cannot  be applied directly and without modification to our protocol, because in these algorithms one is supposed to
know an element and  some power of it; in our case $X\in G$ is known but $X^a$ is hidden  due to  the  conjugation by a secret matrix $T$.\\

{\bf Algorithm 4.1} We propose the following algorithm (an adaptation of  the  Baby Step Giant Step algorithm) for computing the secret keys. Let  $n$ be the order of $X$. So knowing $X$ and $Y=TX^aT^{-1}$,
 we want to compute the secret keys $a$ , $T$  and the exchanged key $TT'X^{ab}T'^{-1}T^{-1}$.
 
1) For $k=1$ to $n$ compute $X^k$, and put the sorted  result in a table.

2) For $x,y\in\F_q$  such that $x^2-y^2\not =0 $ put 
$T_{x,y}=\left(\begin{array}{cc}
x&y\\
y&x
\end{array}\right)$; then
 compute $T_{x,y}YT_{x,y}^{-1}$, and compare it to the table of step (1).

3) If, for some $k_0$ and some $T_{x_0,y_0}$, one has $T_{x_0,y_0}YT_{x_0,y_0}^{-1}=X^{k_0}$, then stop step (2); $a=k_0$ and 
$T=T_{x_0,y_0}^{-1}$  being known, compute
$(T'X^bT'^{-1})^a$ and conjugate it by $T_{x_0,y_0}^{-1}$ to obtain the exchanged secret key.

 As for the complexity of Algorithm 4.1,  we have $O(n)$ group operations in the first step.  Then,  in the second step, we have $O(q^2)$ group operations and $O(nln(n))$ comparaisons.  So, assuming  that a comparaison is much faster than a group operation, we conclude that altogether  the algorithmic cost is  $O(max(n,q^2))$.\\

Taking into account the above  values,  we propose  to take $|\F_q|\simeq 2^{40}$ and  the matrix $X$ of~$GL_2(\F_q[S_3]$ to be of order  $\geq 2^{80}$.\\

 We propose to generate  the invertible matrices $X$  as follows. First, we observe that every matrix  $\left(\begin{array}{cc}
a&c\\
0&b
\end{array}\right)$  with $a$ and $b$  invertible in $\F_q[S_3]$ and no condition on $c$  is invertible, with inverse  $\left(\begin{array}{cc}
a^{-1}&-a^{-1}cb^{-1}\\
0&b^{-1}
\end{array}\right)$. Also  every matrix  $\left(\begin{array}{cc}
a&0\\
c&b
\end{array}\right)$  satisfying  the same conditions  is  invertible,  with inverse 
$\left(\begin{array}{cc}
a^{-1}&0\\
-b^{-1}ca^{-1}&b^{-1}
\end{array}\right)$.
 Then,  we can see that  every matrix of the form 
 $X=\left(\begin{array}{cc}
u&b\\
c&1+cu^{-1}b
\end{array}\right)$  with  $u$ invertible in $\F_q[S_3])$ and no condition on $b,c$  is invertible as well. Indeed, we observe  that $PXQ=Id$ where $P$ and $Q$ are the invertible matrices
$P=\left(\begin{array}{cc}
1&0\\
-cu^{-1}&1
\end{array}\right)$ and $Q=\left(\begin{array}{cc}
u^{-1}&-u^{-1}b\\
0&1
\end{array}\right)$,  leading to 
 $X^{-1}=\left(\begin{array}{cc}
u^{-1}+u^{-1}bcu^{-1}&-u^{-1}b\\
-cu^{-1}&1
\end{array}\right)$. 
By multiplying invertible matrices  of the types above, one can obtain a number of  invertible matrices. \\

 We now determine  $|GL_2(\F_q[S_3])|$, which is helpful for computing the order of elements.\\

{\bf Lemma 4.1.1}:
Suppose the  characteristic  of $\F_q$ is not
 $2$   or   $3$, so that $\F_q[S_3]$ is a semisimple algebra. Then
$|GL_2(\F_q[S_3])|=q^8(q-1)^8(q+1)^4(q^2+1)(q^2+q+1)$.\\

 Proof.
Using the  linear  representations of the  symmetric  group $S_3$ and of the group algebra $F_q[S_3]$, namely
the fact that $S_3$ has three irreducible representations, two of dimension one and the third of dimension two,
one can write $\F_q[S_3]\simeq \F_q\oplus\F_q\oplus Mat_2(\F_q)$ (Wedderburn theorem).  Then we find 
$Mat_2(\F_q[S_3])\simeq  Mat_2(\F_q)\oplus Mat_2(\F_q)\oplus  Mat_2(Mat_2(\F_q) )$,  and 
$GL_2(\F_q[S_3])\simeq GL_2( (\F_q)\oplus GL_2(\F_q)\oplus  GL_4(\F_q)$,  whence 
$|GL_2(\F_q[S_3])|=[q(q-1)^2(q+1)]^2(q^4-1)(q^4-q)q^4-q^2)(q^4-q^3)$,  and 
$|GL_2(\F_q[S_3])|=q^8(q-1)^8(q+1)^4(q^2+1)(q^2+q+1)$.\\

{\large\bf 5. Conclusion}\\

Matrix groups  admit  a natural normal form, making them easy to use for cryptography. Over finite fields special  properties  of matrix groups such
as determinant and eigenvalues can be used to  develop  attacks  against  the protocol  investigated  in this paper. So, in any cryptographic
protocol using matrix groups, one has  first  to verify that the above properties  cannot  be used to defeat the system.
 By using  matrix groups over  a  noncommutative ring such as  the  group algebra of  a  finite group (for  instance  $\F_q[S_n]$),  we can  avoid such attacks.\\

We thank the referee for informing us of some  references.\\
\vfill

{\large\bf References}\\

[1]  D. Boneh, A. Joux, Phong Q. Nguyen, Why textbook Elgamal and RSA encryption are insecure,  Lecture Notes in Computer Science. 1976 (2000), 30-44\\

[2] M.A. Cherepnev, V.M. Sidelnikov, V.V. Yashchenko, Systems of open distribution of keys on the basis of noncommutative semigroups, Russian Acad. Sci.
Dokl. Math. 48 (1994), no. 2, 384-386.\\

[3] P. Dehornoy, Braid-based cryptography, Contemporary Mathematics, 360 (2004),  5-33.\\

[4] I. Gelfand, D. Krob, A. Lascoux, B. Leclerc, V. Retakh, J-Y. Thibon, Noncommutative
symmetric functions, Advances in Math. 112 (1995), no. 2,  218-348.\\

[5]  I. Gelfand, V. Retakh, Determinants of matrices over noncommutative rings, Funct.
Anal. Appl., 25 (1991),no. 2,  91-102.\\

[6]  I.Gelfand, V. Retakh, Quasideterminants \rm 1, Selecta Math. (N.S.) 3 (1997), no. 4, 517-546.\\

[7]  I. Gelfand, S. Gelfand, V. Retakh, R. Wilson, Quasideterminants, Advances in Math., 193 (2005),  56-141.\\

[8]  D. Grigoriev, V. Shpilrain, Authentication from matrix conjugation, Groups, Complexity, Cryptology, 2009, Vol. 1, 199-205.\\

[9] K.H. Ko, S.J. Lee, J.H. Cheon, J.W. Han, J. Kang, C. Park, New public-key cryptosystem using braid groups, Lecture
Notes in Computer Science, 1880 (2000), 166-183.\\

[10] Menezes A.J., Wu Yi-H. The discrete logarithm problem in $GL_n(\F_q)$; ARS Combinatorica. 47 (1997),  23-32.\\

[11] D.N. Moldovyan, N.A.  Moldovyan, A new hard problem over noncommutative finite groups for cryptographic protocols,  Lecture Notes in Computer Science, 6258 (2010), 183-194.\\ 

[12] R.Odonne, D. Varadharajan, P. Sanders, Public key distribution in matrix rings, Electronic Letters, 20  (1984), 386-387.\\

[13] V. Ottaviani, A. Zanoni, M. Regoli, Conjugation as public key agreement protocol in mobile cryptography, SECRYPT, Sci. Te. Press (2010),   411-416.\\

[14] E. Sakalauskas, P. Tvarijonas, A. Raulynaitis, Key agreement protocol using conjugacy and discrete logarithm problems in group representation level,
Informatica, 18  (2007) ,no 1,  115-124.\\

[15] V. Shpilrain, Noncommutative determinants and automorphisms of groups, Comm. Algebra 25 (1997), 559-574.\\

[16] Soojin Cho, Kil Chan Ha, Young-One Kim, Dongho Moon, Key exchange protocol using matrix algebras and its analysis, 
Journal of the Korean Mathematical Society, 42/6  (2005),  1287-1308.

\end{document}